\documentclass{aa}  

\usepackage{graphicx}
\usepackage{txfonts}
\usepackage{hyperref}
\begin{document}

\title{Intermediate-mass stars and the origin of the gas-giant planet-metallicity correlation.}
%The role of the host star.}

\author{J. Maldonado \inst{1}
\and G. M. Mirouh \inst{2}
\and I. Mendigut\'ia \inst{3}
\and B. Montesinos \inst{3}
\and J. L. Gragera-M\'as \inst{3, 4}
\and E. Villaver \inst{5}
}

\institute{INAF - Osservatorio Astronomico di Palermo, Piazza del Parlamento 1, 90134 Palermo, Italy\\
\email{jesus.maldonado@inaf.it}
\and Instituto de Astrof\'isica de Andaluc\'ia (CSIC), Glorieta de la Astronom\'ia s/n, 18008, Granada, Spain
\and Centro de Astrobiolog\'ia (CAB) CSIC-INTA, ESA-ESAC Campus, 28692, Villanueva de la Ca\~{n}ada, Madrid, Spain
\and Departamento de F\'{i}sica de la Tierra y Astrof\'{i}sica, Facultad de Ciencias F\'{i}sicas, Universidad Complutense de Madrid, 28040, Madrid, Spain
\and Instituto de Astrof\'isica de Canarias, 38200 La Laguna, Tenerife, Spain
}

\date{Received September 15, 1996; accepted March 16, 1997}

%\abstract{}{}{}{}{}
% 5 {} token are mandatory

\abstract
% context heading (optional)
% {} leave it empty if necessary  
{
Currently, the number of known planets around intermediate-mass stars (1.5 M$_{\odot}$ < M$_{\star}$ < 3.5 M$_{\odot}$) is rather low and,
as a consequence, models of planet formation have their strongest observational evidence on the chemical signature 
of mostly low-mass (FGK) Main-Sequence (MS) stars with planets.
}
% aims heading (mandatory)
{
We aim to test whether the well-known correlation between the metallicity of the star and the presence of gas-giant planets
found for MS low-mass stars still holds for intermediate-mass stars.
In particular, we aim to understand whether or not the planet-metallicity relation changes as stars evolve from the pre-MS
to the red giant branch. 
}
% methods heading (mandatory)
{
We compile the basic stellar parameters (metallicity, mass and age) of a sample of intermediate-mass stars
at different evolutionary phases %known to host gas-giant planets and
with and without evidence suggesting that they host
%known and without known 
gas-giant planets. The metallicities of the different susbsamples are compared
and set in the context of current models of planet formation and stellar evolution. 
}
% results heading (mandatory)
{
Our results confirm that pre-MS stars with transitional discs with gaps show lower metallicities than pre-MS with flat discs.
We show a tendency of intermediate-mass stars in the MS to follow the gas-giant planet-metallicity correlation,
although the differences in metal content between planet and non-planet hosts are rather modest and
the strength of the correlation is significantly lower than for the less massive FGK MS stars.
For stars in the red giant branch, we find a strong planet-metallicity correlation, compatible with that
found for FGK MS stars.
We discuss how the evolution of the mass in the convective zone of the star's interior might affect the measured metallicity of the star.
In particular, if the planet-metallicity correlation were of primordial origin, one would expect it to be stronger
for less massive stars, as they are longer convective during the stellar evolution.
However, within our sample, we find the opposite. % a stronger planet-metallicity correlation for the more massive stars. 
}
% conclusions heading (optional), leave it empty if necessary
{
The lack of a well-established planet-metallicity correlation in
pre-MS and MS
intermediate-mass stars can be explained by a scenario in which planet formation leads to an accretion of metal-poor material on the surface
of the star. As intermediate-mass stars are mainly radiative the metallicity of the star does not reflect its bulk composition but the composition of the accreted material. 
When the star leaves the MS and develops a sizeable convective envelope, a strong-planet metallicity correlation is recovered. Thus, our results are in line
with core-accretion
%thus, supporting core-accretion 
models of planet formation and the idea that the planet-metallicity correlation reflects a bulk property of the star. 
%We can not completely rule out the formation of planets by disc instability
%5and it is likely that core-accretion and disc instability might simultaneously operate in
%intermediate-mass stars.
}


\maketitle
%
%________________________________________________________________
\section{Introduction}\label{intro}
% ---------------------------------------------------------------

Understanding the origin of stars and planetary systems is one of the major goals
of modern astrophysics. Almost thirty years of exoplanetary science has 
unveiled an astonishing diversity of planetary architectures as well as host stars.
Exoplanets have been discovered not only around solar-like Main-Sequence (MS) stars
%but host stars include
but also around brown dwarfs and low-mass stars, metal-poor stars, giant stars as well as white dwarfs
and pulsars \citep[e.g.][]{2018exha.book.....P}. However, the bulk majority of known exoplanets is still found to orbit around MS stars.
In turn, only about 5\% of known exoplanets
%a small percent of the known exoplanets ($\sim$ 5\%) 
are found around stars with masses between 1.5 and 3.5 M$_{\odot}$,
based on the available data at The Extrasolar Planets Encyclopaedia\footnote{https://exoplanet.eu/home/}
\citep{2011A&A...532A..79S}.

%A quick inspection of the available data at The Extrasolar Planets Encyclopaedia\footnote{https://exoplanet.eu/home/}
%\citep{2011A&A...532A..79S}  reveals that only $\sim$ 5\% of the known exoplanets are found around stars with masses between 1.5 and 3 M$_{\odot}$. 

The role of the host star's chemical composition %(in particular the iron abundance) 
in planet formation
has been largely discussed, with the finding that the frequency of gas-giant planets is a strong
function of the stellar metallicity
\citep[e.g.][]{1997MNRAS.285..403G,2004A&A...415.1153S,2005ApJ...622.1102F}.
Observations of solar-type, FGK dwarfs, MS planet hosts
point towards a metal rich nature of the MS stars %throughout their interiors.
\cite{2005ApJ...622.1102F} showed that the probability of formation
of a gas-giant planet around an FGK-type dwarf depends on the square of the number of iron atoms,
in agreement with the expectation from the
collisional agglomeration of dust grains.
Other explanations invoking the late-stage accretion of metal-depleted material
onto the convective zone of the star \citep{1997MNRAS.285..403G,1997ApJ...491L..51L} were ruled out. 

%Over the range -0.5 < [Fe/H] < +0.5, and 
%For FGK-type MS stars, \cite{2005ApJ...622.1102F} showed that the probability of formation
%of a gas-giant planet depends on the square of the number of iron atoms,
%in agreement with the 
%($P$ < 4 yr, $K$ > 30 ms$^{\rm -1}$) as %around FGK-type MS stars as
%$P(gp)$ =0.03$\times$10$^{2.0[Fe/H]}$. The authors argued that the dependence of $P(gp)$ on the square of the number
%of iron atoms was the natural
%expectation from the
%collisional agglomeration of dust grains. 

Intermediate-mass stars from pre-MS and MS to giants offer the unique 
opportunity of testing how well founded the planet-metallicity relation is, how it changes as the star evolves,
and to unravel its origin.
At the time of writing, no planets have been confirmed around young, intermediate-mass Herbig stars \citep{2023SSRv..219....7B},
although several candidates have been proposed.
For example, an accreting protoplanet has been proposed in the disc around the Herbig star AB Aurigae \citep{2022NatAs...6..751C}
but its detection is the subject of debate \citep{2024AJ....167..172B}. 
Nevertheless, 
high-resolution imaging surveys have shown that many Herbig stars show cavities in their discs, potentially
carved out by giant planets \citep[e.g.][]{2017A&A...603A..21G,2022A&A...658A.112S,2022A&A...667C...1S}.
Based on their spectral energy distributions, the discs around Herbig stars are usually classified as warm, flaring, group I 
and cold, flat, group II \citep{2001A&A...365..476M}. Group I sources have transitional discs with radial cavities or
gaps. 
\cite{2015A&A...582L..10K} proposed that flaring group I sources show a deficit of refractory elements, and thus low values of metallicity.
Recently, \cite{2023A&A...671A.140G} confirmed that group I sources tend to have a lower metallicity than group II sources, as well as
that group I sources tend to have stronger (sub-) mm continuum emission likely related to the presence of giant planets. 
This suggests that giant planets should be frequent around group I/low metallicity Herbig stars. 
According to \cite{2015A&A...582L..10K} and \cite{2018MNRAS.476.4418J}, forming planets in group I sources trap the metal-rich material, while metal-depleted material continues to flow
towards the central star. %Given that intermediate-mass, A- and B-type stars have radiative envelopes, metallicity measurements 
%only reflect the composition of the stellar surface, which is polluted by the metal-depleted material.

Herbig stars will eventually evolve into early-type MS stars \citep[e.g.][]{2023SSRv..219....7B}.
Early-type stars are not optimal targets for radial velocity and transit measurements. Their spectra show
few spectral lines that, in addition, are quite broad and shallow due to the high projected rotation of the star.
Therefore, it is difficult to determine the centroid position of the lines and thus a Doppler shift. 
On the other hand, transit detections are challenged by the expected small transit depths and long transit duration. % {\bf ?}.
As a result the number of known exoplanets around intermediate-mass (M$_{\star}$ $>$ 1.5 M$_{\odot}$) MS is rather low
if compared with the number of planets known around their less massive counterparts.
Previous analysis of the chemical patterns of early-type stars with planets has 
focused in the study of 
a likely relation between peculiar chemical patterns like those of $\lambda$ Boötis 
and metallic-lined A stars, with the presence of giant planets \citep{2021A&A...647A..49S,2022A&A...668A.157S},
finding that there is no a unique chemical pattern for these stars.

The post-MS red giant stars that we observe today are the result of the evolution of early-type MS dwarfs.
Whether the correlation between gas-giant planet occurrence and metallicity extends to giant stars has been
the subject of an intense debate, as several studies have found contradictory results \citep[see e.g.][and references therein]{2013A&A...554A..84M}.
In order to explain the apparent lack of a clear planet-metallicity correlation in giant stars
several interpretations have been put forward. 
These invoke scenarios like 
%Alternative explanations like
the accretion of metal-rich material, higher-mass prototoplanetary discs or the formation of massive
gas-giant planets by gravitational instabilities. % have been put forward to explain these results. 
%{\bf You should mention some more recent result, e.g. Wolthoff+ 2022.} 
In addition, the architecture of planetary systems around evolved stars show some peculiarities with respect to the planets orbiting around MS stars,
such as a lack of close-in planets or higher masses and eccentricities
\citep[e.g.][]{2013A&A...554A..84M}. 

In this work, we aim to unravel the gas-giant planet - stellar metallicity relationship for intermediate mass stars
(1.5 M$_{\odot}$ < M$_{\star}$ < 3.5 M$_{\odot}$) through the formation of stars and planets until
the last stages of their evolution. %This is the goal of this paper, 
%in which we analyse the gas-giant planet-metallicity correlation in a large sample of intermediate-mass stars
%covering all evolutionary phases.
%We believe that the comparison of the metallicity distribution of intermediate-mass (1.5 M$_{\odot}$ < M$_{\star}$ < 3.5 M$_{\odot}$) stars from the pre-MS phase, through the MS, to the post-MS phase
%is needed before an explanation about the nature of the metallicity correlation is invoked. This is the goal of this paper, 
%%in which we analyse the gas-giant planet-metallicity correlation in a large sample of 131 intermediate-mass stars
%covering all evolutionary phases.
The paper is organised as follows: Sect.~\ref{sample} describes the stellar samples analysed in this work and how stellar parameters
are obtained. The metallicity distributions are presented in Sect.~\ref{analysis}.
The results are discussed at length in Sect.~\ref{discussion}. Our conclusions follow in Sect.~\ref{conclusion}.

%________________________________________________________________
\section{Stellar sample}\label{sample}
% --------------------------------------------------------------

Our stellar sample is composed of a total of 131 of stars at different evolutionary stages,
namely the pre-MS phase, the MS, and the red giant phase.
%We had to exclude the sub-giant phase from our analysis as to the very best of our understanding, currently only one subgiant
%planet-host star (HAT-P-67) satisfies our mass criterion. 
%Our lower-mass limit corresponds to the separation between stars with an without a radiative envelope, whilst the upper-mass limit is due to the fact that there are no known planets around more massive stars. 
The subsample of intermediate-mass, pre-MS stars was taken from \cite{2023A&A...671A.140G}. It is composed by 67 Herbig stars with spectral types later than B8 (i.e., T$_{\rm eff}$ below 12000 K).
We discarded those stars without a group I/II SED classification (five stars) as well as the stars with only a lower or upper limit in their metallicities' estimates (11 stars). 
We performed an additional cut and discarded the group II targets more massive
than $\sim$ 3.5 M$_{\odot}$ (six stars). This cut is needed, as no group I star has a mass larger than 3.0 M$_{\odot}$, whilst the discarded group II stars might have masses
as large as $\sim$ 7.0 M$_{\odot}$. 
The selected pre-MS stars cover a range of ages between 2 and 20 Myr, and it is composed of 
20 group I sources and 24 group II sources.
Stellar parameters and group I/II classification were derived from a combination of spectra and photometry
\citep{2021A&A...650A.182G}.
%Basic stellar parameters, namely effective temperature, surface gravity, and metallicity, for the sample of pre-MS stars are derived
%from medium and high resolution spectra and
%are obtained from the comparison of the observed spectra with photospheric models \citep{2021A&A...650A.182G}.
%On the other hand, stellar ages and masses are derived for each star based on optical photometry. %\citep{2021A&A...650A.182G}.
Through this paper we will make the assumption that the gap inferred in the protoplanetary disc of group I stars is due to the presence of (at least)
one giant-planet \citep{2023A&A...671A.140G}.
In the same line, group II stars will form the corresponding comparison sample, that is, the sample of pre-MS stars without known planets.

The subsample of MS intermediate-mass stars with and without known planets is drawn from \cite{2021A&A...647A..49S,2022A&A...668A.157S}.
It is composed by a total of 28 stars (13 planet hosts; 15 stars without known gas-giant companions).
The stellar mass varies between 1.5 and 2.2 M$_{\odot}$, whilst the stars have ages ranging from 144 Myr to 1.5 Gyr. 
Stellar parameters are determined in a homogeneous way from high-resolution spectra \citep{2021A&A...647A..49S,2022A&A...668A.157S}. 

The subsample of post-MS, giant stars mainly comes from \cite{2013A&A...554A..84M} and \cite{2016A&A...588A..98M} to which
we added six additional planet-hosts not previously analysed. 
The selected giant sample covers a range of stellar masses between 1.5 and 3.6 M$_{\odot}$ and ages between
240 Myr and 3 Gyr. It is composed of 21 planet hosts and 38 comparison stars (that is, stars without known companions). 
Stellar parameters were derived from high resolution spectroscopy and photometry \citep{2013A&A...554A..84M}.

%Stellar parameters of these stars are derived from the measured equivalent width of spectral lines in high-resolution optical spectra
%by applying the iron ionisation, match of the curve of growth, and excitation equilibrium conditions
%\citep[see e.g.][]{2013A&A...554A..84M}. Stellar masses and ages are computed from photometry and parallaxes
%%Evolutionary parameters are computed from {\sc Hipparcos} V magnitudes and parallaxes
%throughout a Bayesian framework
%\citep{2006A&A...458..609D}.

%Figure~\ref{diagrama_hr} shows the Hertzsprung-Russell (HR) diagram of the stars studied in this work. 
%The main properties of the sample are provided in Appendix~\ref{app_tables}.

% --------============================--------------
% Fig. 1. HR diagram
% ----------------------++++++++++++++++++++++++++++
\begin{figure}[htb]
\centering
\includegraphics[scale=0.60]{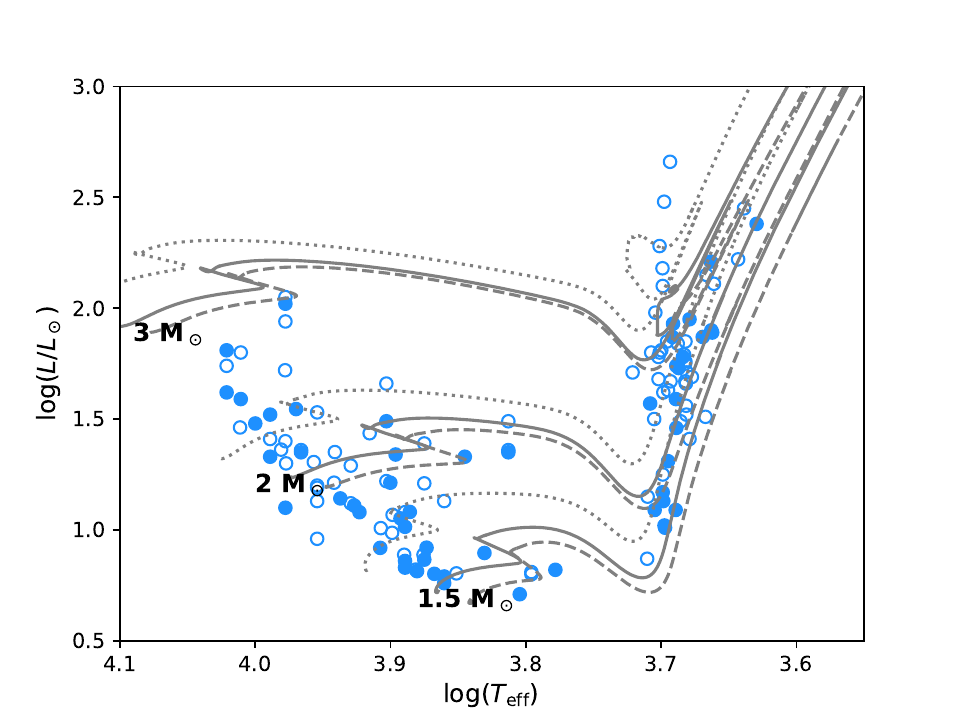}
\caption{
Luminosity versus T$_{\rm eff}$ diagram for the observed stars.
Planet hosts are plotted with filled symbols.
Evolutionary tracks computed with MESA are overplotted (see text). For each mass, three tracks are plotted, corresponding to [Fe/H] = -0.40 (Z = 0.0057, dotted lines), [Fe/H] = + 0.00 (Z = 0.0142, solid lines), and [Fe/H] = +0.20 (Z = 0.0225, dashed lines).
}
\label{diagrama_hr}
\end{figure}

Figure~\ref{diagrama_hr} shows the Hertzsprung-Russell (HR) diagram of the stars studied in this work,
whilst the main properties of the sample are provided in Appendix~\ref{app_tables}.
Given that the planet-metallicity correlation applies only to gas-giant planets \citep[e.g.][]{2010ApJ...720.1290G,2011A&A...533A.141S},
before we proceed with the analysis, we carefully checked that all the planets in the MS and giant star subsamples 
have (minimum)-masses larger than 30 M$_{\oplus}$, which is also the % type of planets
type of planets that should be expected around group I pre-MS stars.

%________________________________________________________________
\section{Analysis}\label{analysis}
%----------------------------------------------------------------
%\subsection{Metallicity distribution}\label{analysis_ecdf}  

\begin{figure*}[htb]
\centering
\includegraphics[scale=0.55]{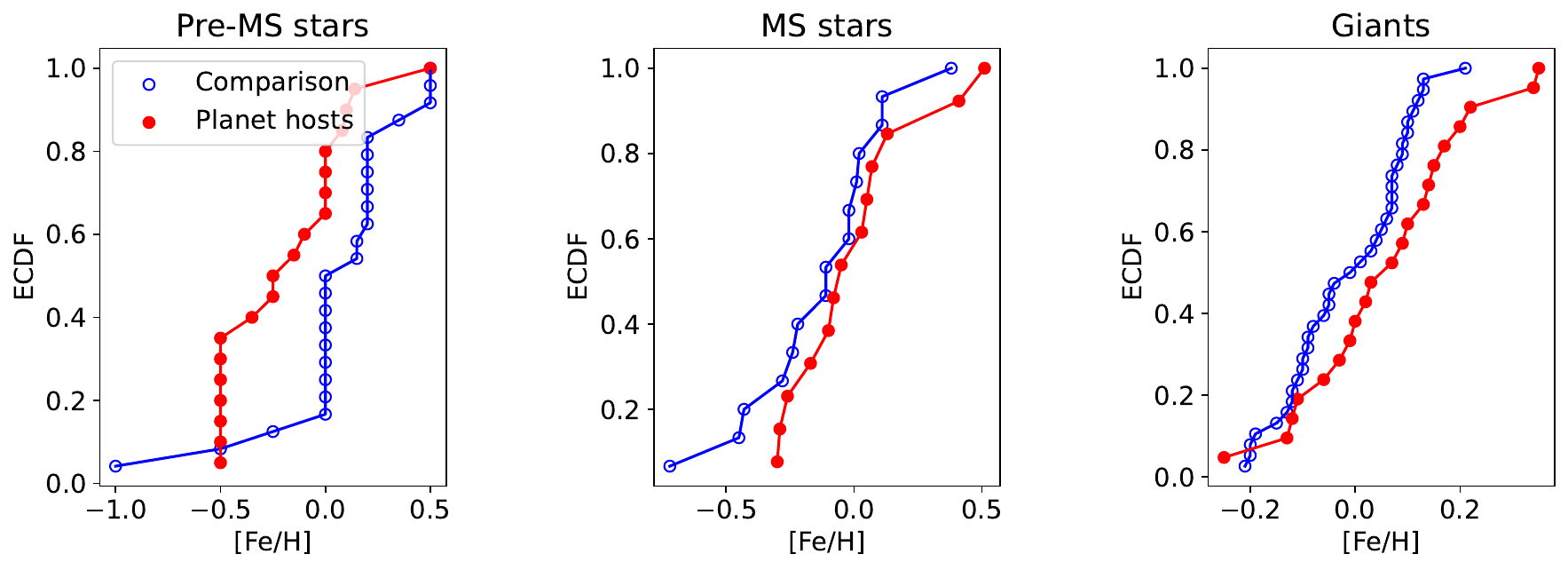}
\caption{
Comparison of the [Fe/H] empirical cumulative distribution function (ECDF) between planet hosts (red circles) and stars without known planets
(blue circles) for the different subsamples studied in this work, from pre-MS stars to  giants.}
\label{ecdf_subsamples}
\end{figure*}

The cumulative distribution function of the metallicity for the different subsamples
analysed in this work 
is presented in Fig.~\ref{ecdf_subsamples}.
For guidance some statistical diagnostics are also provided in Table~\ref{ks_table}.
To assess whether the metallicity distribution of both comparison and planet-hosts are equal from a statistical point of view,
for each subsample, we performed two different statistical tests: a standard two-sample Kolmogorov-Smirnov (K-S)
\citep[e.g.][]{1983MNRAS.202..615P}
and the Anderson-Darling (A-D) \citep[e.g.][]{5f1b3ae3-6190-3984-ae67-e75177b0a848} test. 

There are a few interesting facts to be derived from these distributions and their statistical tests.
Group I pre-MS stars tend to have a lower metallicity distribution than group II pre-MS stars. 
The K-S test confirms that both distributions differ within a confidence level of 99.9\%,
in agreement with the A-D test that shows that 
the null hypothesis that the comparison and the planet host metallicity distribution are similar 
(come from the same parent population)
can be rejected at the
0.1\% level as the returned test value (6.76) is  greater than the critical value for 0.1\% (6.55).
%This results are in agreement with the work of \cite{2023A&A...671A.140G}
If we take the assumption that the disc structures found on group I sources are likely related to the presence of
gas-giant planets, that means that pre-MS planet-hosts do not show metal enrichment, but rather a deficit of metals
with respect to their respective comparison sample.
This result confirms the previous findings of \cite{2015A&A...582L..10K} and \cite{2023A&A...671A.140G}.

For intermediate-mass MS stars there seems to be a tendency of planet hosts to have slightly larger metallicity values
than their corresponding comparison sample. However, none of the performed statistical tests confirms this fact. %assumption. 
The K-S test provides a probability that both samples (planet hosts and comparison stars) share a similar metallicity
distribution of roughly 60\%, whilst from the A-D we just simply cannot reject the 
null hypothesis that the comparison and the planet host metallicity distribution are similar.

We find that the gas-giant planet - stellar metallicity correlation is also present in intermediate-mass giant stars
in agreement with recent works \citep[e.g.][]{2013A&A...554A..84M,2015A&A...574A.116R,2022A&A...661A..63W}.
The A-D test shows that the metallicity distribution of planet hosts and comparison stars differ
at the 5\% level although not at the 2.5\% level. 
%As we will further discuss, this fact points towards a primordial origin of the planet-metallicity correlation. 

% ----------------------
% Fig. 1. ECDF
% ----------------------

% ----------------------
% Table. Results
% ----------------------
\begin{table*}
\centering
\caption{Comparison between the [Fe/H] of the different subsamples.
For each subsample we list the number of stars ($N$) as well as the mean and the standard deviation values of
the [Fe/H] distribution. For the statistical tests, we provide % $D$, which denotes
the test statistic value, $D$, as well as the asymptotic $p$-value.
The A-D test critical values for significance levels of 5\%, 2.5\%, 1\%, 0.5\%, and 0.1\% are
1.96, 2.72, 3.75, 4.59, and 6.55, respectively. Values of $\alpha$, defined as in Eq.~\ref{metaleq}, are provided for the bin fitting as well
as for the Bayesian fit.}
\label{ks_table}
\begin{tabular}{lcccccccccccc}
\hline
Sample  & \multicolumn{3}{c}{Comparison}   &  \multicolumn{3}{c}{Planet hosts}      & \multicolumn{2}{c}{K-S test} & \multicolumn{2}{c}{A-D test} & $\alpha$         & $\alpha$    \\ 
&  $N$  &Mean       &   $\sigma$    & $N$   &  Mean   &       $\sigma$              &  $D$          & $p$-value    &  $D$          & $p$-value    & Bin fitting      & Bayesian fit \\
\hline
pre-MS &   24 &      0.07 &        0.31   & 20  &  -0.20  &       0.28              &  0.47  &    0.001            &  6.76         &  0.002       & -0.29 $\pm$ 0.13 &  -1.40 $\pm$  0.83  \\
MS     &   15 &     -0.11 &        0.26   & 13  &  -0.05  &       0.25              &  0.26  &    0.610            & -0.21         &  0.456       &  0.50 $\pm$ 0.19 &   0.80 $\pm$  0.60  \\
Giants &   38 &      0.00 &        0.11   & 21  &   0.07  &       0.15              &  0.31  &    0.121            &  2.31         &  0.040       &  1.64 $\pm$ 0.42 &   1.77 $\pm$  0.85  \\ 
\hline
\end{tabular}
\end{table*}

%# PMC evolutionary stages stars with/without planets
%# 2024-07-08 11:26:06.870780
%# n_comp comp_mean comp_st  n_planets planets_mean planet_std  D-value p-value n_eff
%# Pre-MS stars ++++++++++++++++++++++++++++++++++++++
%# -----------------------------------------------------
%24         0.07         0.31 20        -0.20         0.28         0.47      0.00956         10.9
%Anderson-Darling: statistic, p-value, critical values 25%, 10%, 5%, 2.5%, 1%, 0.5%, 0.1%
%6.756252000867449 0.0024 [0.325 1.226 1.961 2.718 3.752 4.592 6.546]
%# MS stars ++++++++++++++++++++++++++++++++++++++
%# -----------------------------------------------------
%15        -0.11         0.26 13        -0.05         0.24         0.26      0.60897          7.0
%Anderson-Darling: statistic, p-value, critical values 25%, 10%, 5%, 2.5%, 1%, 0.5%, 0.1%
%-0.20934976091341256 0.4561 [0.325 1.226 1.961 2.718 3.752 4.592 6.546]
%# Giants ++++++++++++++++++++++++++++++++++++++
%# -----------------------------------------------------
%38         0.00         0.11 21         0.07         0.15         0.31      0.12112         13.5
%Anderson-Darling: statistic, p-value, critical values 25%, 10%, 5%, 2.5%, 1%, 0.5%, 0.1%
%2.3095194828030317 0.0398 [0.325 1.226 1.961 2.718 3.752 4.592 6.546]

\begin{figure*}[!htb]
	\centering
	\includegraphics[scale=0.60]{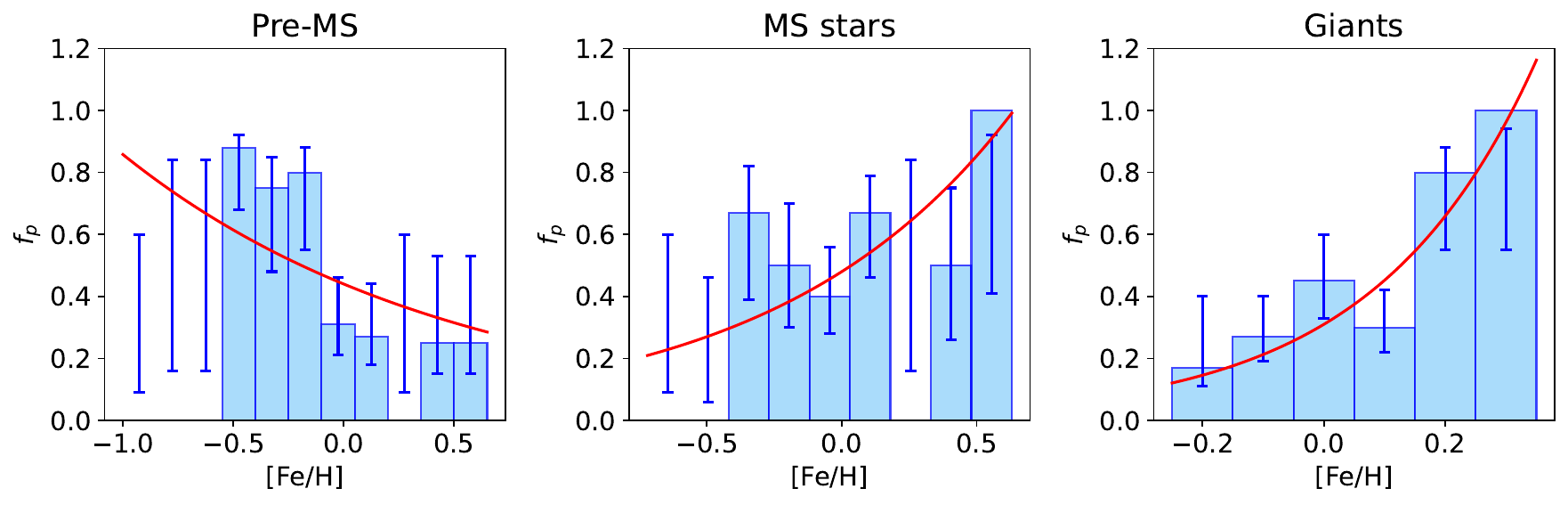}\\
	\caption{Frequency of gas-giant planets as a function of the stellar [Fe/H]
	for each of the subsamples analysed in this work.
	The best bin fitting is shown in red.
	Vertical lines show the range in planetary fraction that delimits 68.2\%
	of the integrated probability function (see Eq.~\ref{bin_eq}).
	}
	\label{bin_fit}
\end{figure*}

Figure~\ref{bin_fit} shows the frequency of gas-giant planets
as a function of the stellar metallicity for the different subsamples.
That is, for each metallicity bin, the number of known gas-giant planets is divided by 
the total number of stars of the bin. 
The uncertainties in the frequency of each bin are calculated using the binomial distribution
(each star either has or not has a planet),

\begin{equation}\label{bin_eq}
	P(f_{p},n,N) = \frac{N!}{n!(N-n)!}f^{n}_{p}(1-f_{p})^{N-n}
\end{equation}

\noindent where $P(f_{p},n,N)$ provides the probability of $n$ detections given a sample
of size $N$ when the true planetary companion frequency is $f_{p}$. 
We follow the common practice of reporting the range in planetary fraction that delimits 68.2\% 
of the integrated probability function, which is equivalent to the 1$\sigma$ limits for a Gaussian
distribution \citep[see e.g.][]{2003ApJ...586..512B,2006ApJ...649..436E}.

Following previous works \citep{2005ApJ...622.1102F,2007ARA&A..45..397U}, the fraction of stars with planets was fitted to a function
with a function of the form,

\begin{equation}\label{metaleq} 
f (\rm{[Fe/H]})  = C \times 10^{\alpha{\rm[Fe/H]}}
\end{equation}

The derived $\alpha$ values are provided in Table~\ref{ks_table}.
For pre-MS stars, we see that
the distribution peaks at metallicities around -0.25,
and then, for higher metallicity values, it remains flat. As a consequence, the derived $\alpha$ value is slightly negative $\sim$ {\mbox -0.29}.
MS and giant stars show the well-known planet metallicity correlation, although with some differences.
For intermediate-mass MS the strength of the correlation, $\alpha$ is only $\sim$ 0.5 which is significantly lower than the value found
for their less massive FGK MS counterparts, $\alpha$ $\sim$ 2 \citep{2005ApJ...622.1102F}.
On the other hand, for giants, we found a strong planet-metallicity correlation with a parameter $\alpha$ $\sim$ 1.6 which is compatible %within errors
with the correlation found for FGK-type MS stars.

% -----------------------------------
%\subsection{Mass effect}
% -----------------------------------

% It is also known that the planet occurrence rate, in addition to the metal content of the star, does also rise monotonically with
% the stellar mass \citep[e.g.][]{2010PASP..122..905J,2020A&A...644A..68M}.
% The bottom panel of Fig.~\ref{bin_fit} shows the fraction of gas-giant planets as a function of the stellar mass. In this case,
% the fraction of stars with planets was fitted to a power-law function $f =CM_{\star}^{\alpha}$.
% We found that all samples, irrespective or their evolutionary status, show a rather flat distribution without a clear mass-planet dependency
% that translate into a negative $\alpha$ value but with large uncertainties. 
% This is likely an effect due to the rather small range of masses covered by the different subsamples studied in this work.

% >>>>>>>>>>>>>>>>>>>>>>>>>>>>>>>>>>>>>>>>>>>>>>>>>>>>
% Figure 2
% <<<<<<<<<<<<<<<<<<<<<<<<<<<<<<<<<<<<<<<<<<<<<<<<<<<<

%-----------------------------------------------------------------
%\subsection{Bayesian analysis}
% -----------------------------------------------------------------
In order to further test the planetary frequency as a function of the stellar metallicity, we use
a Bayesian approach.
Details of the analysis can be found in e.g. \citet{2010PASP..122..905J,2020A&A...644A..68M}.
In brief, the planetary frequency is related to metallicity through
Eq.~\ref{metaleq}, with $X$ = (C, $\alpha$) being the parameters to be optimised.
We model the data as a series of Bernoulli trials. The probability of a specific model $X$,
considering our data $d$ is given by the Bayes theorem:

\begin{equation}
P(X|d)\propto P(X)\prod_{i}^{H}f([Fe/H]_{i})\times\prod_{j}^{T-H}[1-f([Fe/H]_{j})]
\end{equation}

\noindent where T is the total number of stars and H is the number of planet hosts.
The likelihood is then given by:

\begin{eqnarray} 
\mathcal{L}\equiv\log P(X/d)\propto\sum_{i}^{H}\log f([Fe/H]_{i})+ \sum_{j}^{T-H}\log[1-f([Fe/H]_{j})] \nonumber \\
+\log P(X)
\end{eqnarray} 

In order to account for uncertainties in the metallicity determination, we
assume a Gaussian probability distribution with mean ($[Fe/H]_{i}$) and
standard deviation ($\sigma_{[Fe/H]i}$). In this way, the predicted planet fraction
for the $i$th star is:

\begin{equation}
f([Fe/H]_{i})=\int p_{obs}([Fe/H]_{i})f([Fe/H])d[Fe/H]
\end{equation}

Our results are shown in Table~\ref{ks_table} where the mean $\alpha$ values and their corresponding
uncertainties are provided, while Fig.~\ref{corner_plots} shows the marginal posterior
probability.
The Bayesian analysis confirms the results from the bin fitting. That is, 
intermediate-mass 
MS planet hosts show a weaker planet-metallicity correlation than
giant stars, while pre-MS stars show an anti-correlation.

%________________________________________________________________

\section{Discussion}\label{discussion} 
%----------------------------------------------------------------

 In this section we discuss our results in the framework of current planet formation models.
 The two main formation models for giant planets are the the core accretion and the disc instability.
 The core accretion model starts with the formation of a massive planet core, followed by a rapid accumulation of a massive gas envelope.
 Within this framework, a high metallicity environment implies a high dust-to-gas-ratio in the protoplanetary disc that facilitates condensation,
 and accelerates accretion before the gas disc is lost \citep[e.g.][]{1996Icar..124...62P,2003ApJ...598L..55R,2004A&A...417L..25A,2012A&A...541A..97M}.
 In the gravitational disc instability, giant planets form by the contraction of gaseous condensations in a massive self-gravitating disc
 \citep[e.g.][]{2002ApJ...567L.149B,2017ApJ...836...53B}.
 Although initial disc instability models are not able to explain the correlation of giant planet occurrence and stellar metallicity,
 the inclusion of pebble accretion into the models might overcome this difficulty 
 \citep{2015MNRAS.448L..25N}.

% -----------------------------------------------
\subsection{Core accretion}
% -----------------------------------------------

As pointed out in Sect.~\ref{analysis} we find that the metallicity distribution of pre-MS Herbig group I stars
is shifted towards lower values with respect to the one of its corresponding group II comparison sample, in agreement with previous claims \citep{2023A&A...671A.140G}.
Assuming that the cavities found in group I sources are carved out by the presence of gas-giant planets, this seems to imply
that pre-MS stars with planets do not follow the gas-giant planet-metallicity correlation.
Indeed, our analysis points out that the frequency of gas-giant planets around pre-MS star does not depend on the stellar metallicity 
or even might show a anticorrelation. % (with a $\alpha$ parameter $\sim$ {\mbox -0.63}).
Although this result apparently contradicts the gas giant planet-metallicity correlation, it may
be explained within the theoretical framework by 
\cite{2015A&A...582L..10K} and \cite{2018MNRAS.476.4418J}. 
%suggest an explanation to a this
%result which apparently, contradicts the gas-giant planet-metallicity correlation. 
According to these authors,
forming planets in protoplanetary discs block the accretion of part of the dust, while gas continues to flow towards the star.
Given that Herbig stars have radiative envelopes, the mixing timescales with the interior are large, of the order of several Myr.
Thus, metallicity measurements reflect the metal composition of the accreted metal-poor material that pollutes the external layers of the star. 
Numerical simulations of the evolution of pre-MS stars have shown that stars with
T$_{\rm eff}$ > 7000 K may show large metallicity deficits, by 0.6 dex or more, in the presence of efficient planet formation
\citep{2018A&A...618A.132K}.

Herbig stars evolve into early-type MS stars within a characteristic time scale of a few Myr.
Typical timescales are of the order of $\sim$ 3 Myr for a 3 M$_{\odot}$ star and 
$\sim$ 20 Myr for a 1.5 M$_{\odot}$ star.
Therefore, slow non-convective processes \citep[see e.g.][]{2018MNRAS.476.4418J} might have partially mixed the metal-poor accreted material during the pre-MS phase with the interior.
Thus, under the assumption that gas-giant planets form in metal-rich environments,
the question is whether or not 
the non-convective processes in intermediate mass MS stars are able to reset the original metal-rich composition of the star.
Our results suggest a rather weak planet metallicity correlation with $\alpha$ $\sim$ 0.8.
This implies that, if the planet-metallicity correlation had a primordial origin, then, the slow mixing processes in stars with a radiative surface % stars
are able to partially recover the bulk composition of the star.
However, we should caution that this interpretation needs of a robust statistical confirmation. As we have seen in Sect.~\ref{analysis}, we cannot rule out
the possibility that planet hosts show a similar metallicity distribution that stars without known planets.

 In the hypothesis that the observed correlation between the metallicity of the star and the presence of gas-giant hosts is a bulk property of the star, 
 it should also holds for red giants that are convective for the most part.
% The reason is that such stars become convective for the most part. % and have had enough time to transfer to the surfaces the primordial metallicity hidden in their stelar interiors.
 %having left the MS when they exhaust the hydrogen in the core, have larger radii, cooler photospheres,
 %and are convective for the most part. 
 As shown before, we found a strong planet-metallicity correlation 
 for red giants that is is compatible with the correlation found for low-mass FGK MS stars.
 Thus, our results support the idea that gas-giant planets are more easily formed in high metallicity environments. 
% Within this framework, a high metallicity environment implies a high dust-to-gas-ratio in the protoplanetary disc that facilitates condensation,
% and accelerates accretion before the gas disc is lost \citep{1996Icar..124...62P}.

% -----------------------------------------------
\subsection{Planet-metallicity correlation and the role of convective envelopes.}
% -----------------------------------------------

\begin{figure}[!htb]
\centering
\includegraphics[scale=0.6]{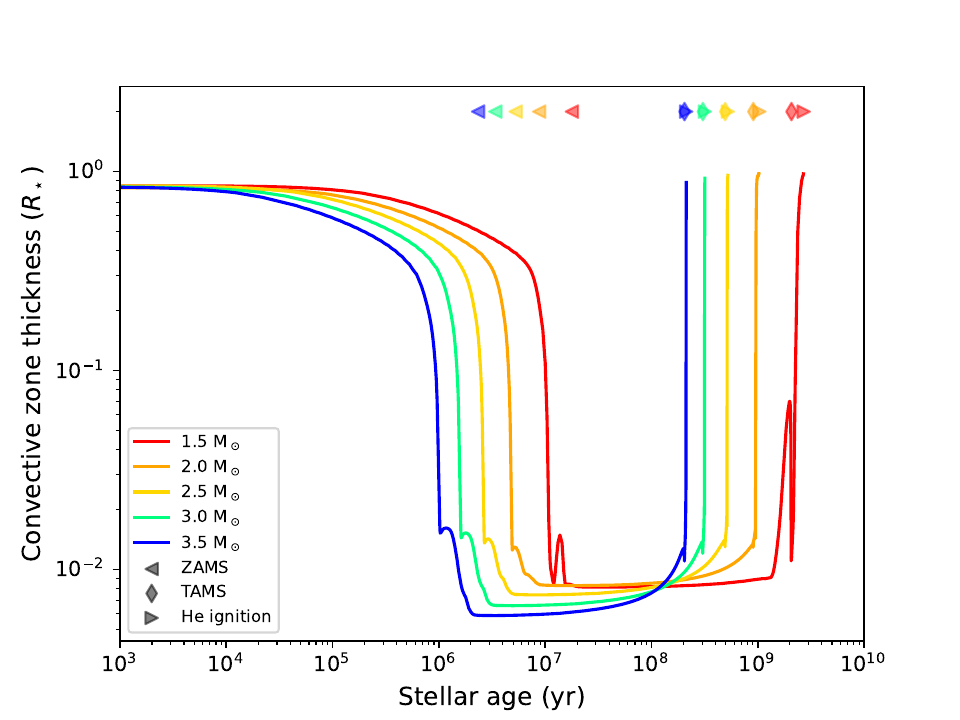}
\includegraphics[scale=0.6]{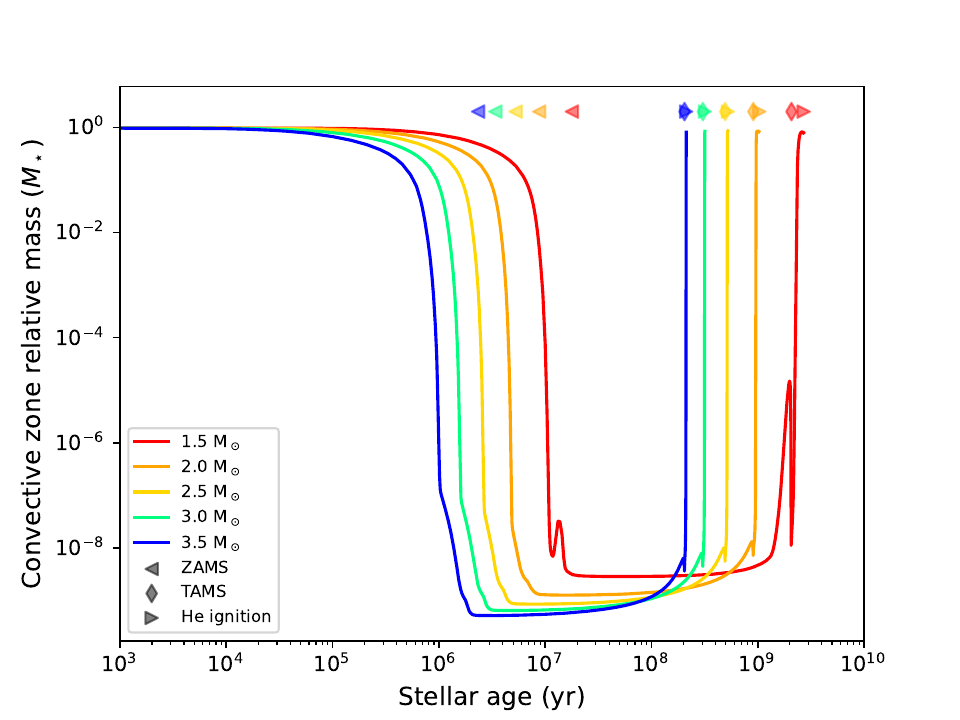}
\caption{
Top panel: thickness of the surface convection zone, bottom panel: mass included in said convection zone (zero means no surface convection and one means an
entirely convective star). The symbols mark the zero-age main-sequence (leftward triangle), the terminal-age main sequence (diamond) and helium ignition (rightward triangle). Different colours indicate different stellar masses.
}
\label{mesa_evol}
\end{figure}

In order to unravel the possible role of the convective zone in the gas-giant planet-metallicity correlation,
we compute a series
of evolutionary tracks by using the Modules for Experiments in Stellar Astrophysics (MESA) package 
\citep{2011ApJS..192....3P,2013ApJS..208....4P,2015ApJS..220...15P,2018ApJS..234...34P,2019ApJS..243...10P,2023ApJS..265...15J}.
Tracks were computed for stellar masses between 1.5 M$_{\odot}$ and 3.5 M$_{\odot}$ by using the solar composition (Y=0.2703, Z=0.0142)
from \citet{2009ARA&A..47..481A} and a mixing length parameter of 1.82. The simulations do not include overshooting, semi-convection or mass-loss processes.
Figure~\ref{mesa_evol} shows the evolution of the thickness and mass included in the convection zone 
of intermediate-mass stars. % in the pre-MS (top), the MS (middle), and in the red giant phase (bottom).

Newly born, pre-MS stars are fully convective. However, they develop a radiative envelope in a short timescale.
This timescale is mass-dependent and more massive stars (blue line) turn on radiative more rapidly.
%For example, a 1.5 M$_{\odot}$ turns radiative in $\sim$ {\bf 50000} yr, whilst a 3.5 M$_{\odot}$ only takes %needs
%$\sim$ thousand years to become radiative.
%These numbers are in agreement with previous estimates \citep{2018A&A...618A.132K}.

For MS stars, it can be seen that irrespective of their masses, they remain with a radiative envelope during their MS evolution. 
While a tendency towards slightly larger convective zones during their evolution can be seen, the
mass in the convective zone remains always below $\sim$ 10$^{\rm -5}$ the total mass of the star.

Finally, stars at the beginning of the Hertzsprung gap and the red giant phase have a radiative envelope.
They develop a convective zone 
during their evolution across the Hertzsprung gap/subgiant phase.
%the red giant branch. 
The models show that less-massive stars (red line)
develop a convective envelope
more rapidly than the more-massive stars (blue line).

% >>>>>>>>>>>>>>>>>>>>>>>>>>>>>>>>>>>>>>>>>>>>>>>>>>>>
% Figure MESA evolution
% <<<<<<<<<<<<<<<<<<<<<<<<<<<<<<<<<<<<<<<<<<<<<<<<<<<<

\begin{figure*}[!htb]
\centering
\begin{minipage}{0.33\linewidth}
\includegraphics[scale=0.55]{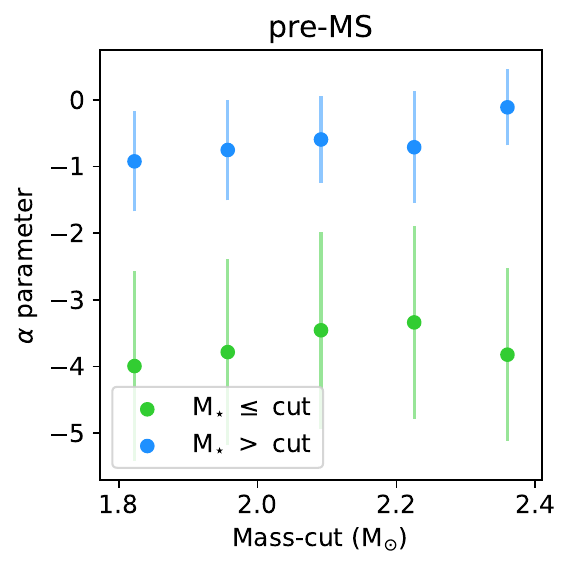}
\end{minipage}
\begin{minipage}{0.33\linewidth}
\includegraphics[scale=0.55]{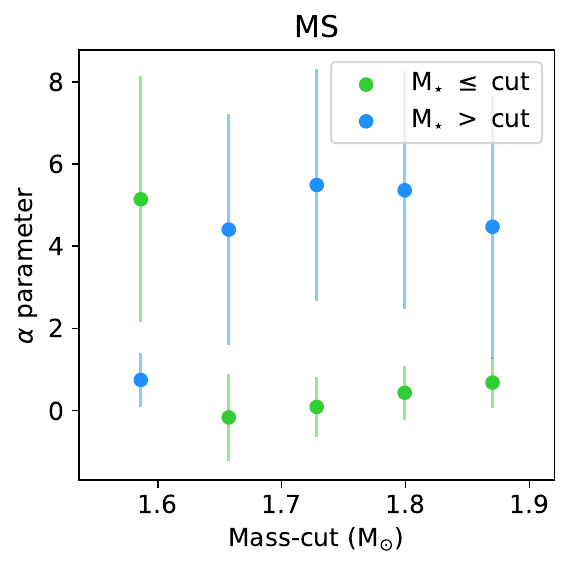}
\end{minipage}
\begin{minipage}{0.33\linewidth}
\includegraphics[scale=0.55]{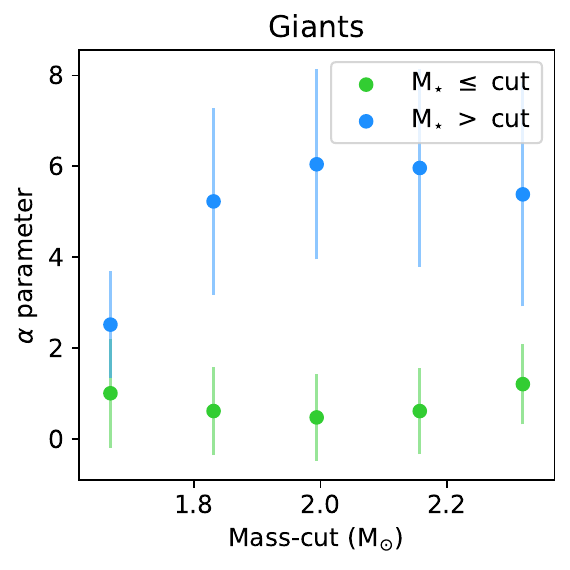} %MESA_evol_giants.pdf}
\end{minipage}
\caption{
$\alpha$ values derived for the ``less massive''(green) and 
``more massive'' (blue) subsamples as a function of the adopted mass cut-off. 
}
\label{pmc_mass_cut}
\end{figure*}

These results imply that stellar evolution plays a fundamental role in our understanding of the gas-giant planet-metallicity correlation. 
In summary, models predict that high-mass stars with planets should be longer radiative, both in the pre-MS as well as in the red giant phase,
and therefore are likely to be more
polluted by the accretion of metal-poor material than low-mass stars.
In other words, the stellar surfaces of lower mass stars should better reflect primordial compositions because mixing by convection
lasts longer for these stars.
Therefore, if we assume that the observed planet-metallicity correlation has a primordial origin, we might expect it to be stronger 
for the less massive stars. 
In order to test whether this effect is or not present in our sample, we divide the different samples % of pre-MS and giant stars 
into ``less massive'' and
``more massive'' by imposing a mass cut-off.
In order to avoid using a somehow arbitrary cut-off and/or small subsamples size,
five different values of mass were considered for the cut. 
%in 2.1 M$_{\odot}$. This cut is somehow arbitrary, but it divides both samples in a roughly equal-size subsamples.
%We exclude the MS sample from this exercise, as we have seen that irrespective of their masses, these stars retain a radiative envelope during their 
%MS evolution. 
The parameter $\alpha$ (defined in Eq.~\ref{metaleq}) was taken as  a measure of the ``strength'' of the planet-metallicity correlation
and a Bayesian analysis was performed for each subsample. The results are
shown in Fig.~\ref{pmc_mass_cut} where the derived $\alpha$ values are shown for the less-massive (green symbols) and more-massive (blue colour) subsamples as a function of the
mass value used for the cut-off. 

% >>>>>>>>>>>>>>>>>>>>>>>>>>>>>>>>>>>>>>>>>>>>>>>>>>>>
% Figure: Alpha evolution as a function of mass
% <<<<<<<<<<<<<<<<<<<<<<<<<<<<<<<<<<<<<<<<<<<<<<<<<<<<

For pre-MS stars we obtain a somehow stronger planet-metallicity correlation for the more massive stars, in disagreement with our expectations.
We note that the
derived $\alpha$ values are not dependent on the mass used to define the high- and low-mass subsamples.
On the other hand, when considering the stars in the MS and in the red giant phase,
we also find that the planet-metallicity correlation is clearly stronger for the more-massive stars
that remain radiative longer than their less-massive counterparts.
Again, this result contradicts our expectation.
Furthermore, we find a strong dependency of $\alpha$ on the selected cut-off mass,
which can take values as high as six when considering a mass cut-off $\gtrsim$ 2.0 M$_{\odot}$.

%given in Table~\ref{fit_table_mass}.
%For pre-MS stars we derive $\alpha$ = -1.33
%for the less massive stars, whilst
%for the more massive pre-MS stars we obtain $\alpha$ = -0.43. 
%That is, we obtain a somehow stronger planet-metallicity correlation for the more massive stars, in disagreement with our expectations.
%On the other hand, when considering the stars in the red giant phase, we find $\alpha$ = 0.89 and $\alpha$ = 6.90, for the less- and the more-
%massive, respectively. Again, this result contradicts our expectation as the planet-metallicity correlation is clearly stronger for the more-massive stars
%that remain radiative longer than their less-massive counterparts. 

% ----------------------
% Table. Results
% ----------------------
%\begin{table}
%\centering
%\caption{Parameters of the bayesian fit when a mass cut-off of 2.10 M$_{\odot}$ is considered. Values corresponding to the 68.2\% confidence interval are provided.}
%\label{fit_table_mass}
%\begin{tabular}{lc}
%\hline
%Sample &   $\alpha$                                \\
%\hline
%pre-MS with M$_{\star}$ < 2.1 M$_{\odot}$  & -1.33 (-2.10, -0.49) \\
%pre-MS with M$_{\star}$ > 2.1 M$_{\odot}$  & -0.43 (-0.85,  0.01) \\
%\hline
%MS with M$_{\star}$ < 2.1 M$_{\odot}$  & \\ 
%MS with M$_{\star}$ > 2.1 M$_{\odot}$  & \\
%\hline 
%Giants with M$_{\star}$ < 2.1 M$_{\odot}$  & 0.89 (       -0.04,        1.92) \\
%Giants with M$_{\star}$ > 2.1 M$_{\odot}$  & 6.90 (        4.83,        9.48) \\    
%\hline
%\end{tabular}
%\end{table}

% -------------------------------------------------
\subsection{Caveats and possible scenarios}  
% -------------------------------------------------

The previous contradictory results lead us to the question of whether or not some biases might affect our results.
To start with, we note that the sample sizes are reduced significantly when applying the mass cut-off. 
Second, even if the formal errors in the stellar masses are rather low, the accurate determination of stellar masses for pre-MS and giants stars is
rather difficult as the evolutionary tracks are very close and small changes in temperature, luminosity or stellar metallicity
might be of importance. 
Furthermore, it could be the case that the difference in mass between the less- and more- massive stars is not large enough
to find a statistically significant difference. For example, if we set the mass cut-off at 2.1 M$_{\odot}$ then, the mean mass of the less-massive pre-MS is 1.8 M$_{\odot}$ while 
the mean mass of the more-massive pre-MS is 2.5 M$_{\odot}$ (similarly, for giants we have mean masses of 1.7 M$_{\odot}$ and 2.6 M$_{\odot}$ for the
less- and more- massive stars, respectively). Furthermore, we caution that the uncertainties in the derived $\alpha$ values are rather large. 
%In addition, chemical abundances can be affected by individual processes {\bf do not like individual processes, any better word?}.
%rotation, binarity + episodes of planet engulfment. 

%Finally, it should be noted that it is well known that the planet occurrence rate, in addition to the metal content of the star,
%does also rise monotonically with the stellar mass \citep[e.g.][]{2010PASP..122..905J}. However, for all the samples studied here (pre-MS, MS, and red giants)
%we find a rather flat distribution without any clear mass-planet dependency.

% -------------------------------------------------
%\subsection{Planet engulfment} % and iron abundances}
% -------------------------------------------------
In addition, planet engulfment can produce refractory element enhancements within
the engulfing star convective region due to ingestion of rocky planetary material \citep[e.g.][]{2018ApJ...854..138O}.
In a recent work, \cite{2024A&A...686L...1M} shows that
stellar magnetospheres act as a protoplanetary disc barrier
preventing unlimited planet migration. %Magnetospheres disappear in Herbig stars with masses $\gtrsim$ 3-4 M$_{\odot}$
%\citep{2022ApJ...930...39V} thus, favouring planet-engulfment. 
As discussed in that work, magnetospheres become smaller as the stellar mass increases, potentially disappearing in Herbig stars with masses
$\gtrsim$ 3-4 M$_{\odot}$
\citep{2020MNRAS.493..234W,2022ApJ...930...39V}.
Thus, planet engulfment by intermediate-mass host stars is more likely as the stellar mass increases, which is in line with the mass dependence of the planet metallicity
correlation we found for the pre-MS and MS samples.
Along this line, \citet{2025A&A...693A..47S} show that engulfed planets have a higher amount of refractories and that
engulfment is more likely to occur in systems that come from more massive and more metal-rich protoplanetary discs.

On the other hand, tidal interactions in the star system as the star evolves off
the MS can lead to variations in the planetary orbits as well as to the engulfment of
close-in planets \citep{2009ApJ...705L..81V}. 
The planet accretion process might lead to a transfer of angular momentum to the 
stellar envelope, which ultimately can spin up the star and even modify its
chemical abundances.
This would be in line with the enhancement of the planet-metallicity correlation we find during the post-MS phase.

As we have discussed, more massive stars retain longer their radiative envelopes.
Hence, contamination by planet engulfment is expected to be more important. 
Therefore, the question on whether or not planet engulfment may affect the metallicity content of
more massive stars arises. An enrichment in refractory elements produced by planet engulfment 
may be a suitable explanation for the strong planet metallicity correlation found for the more
massive stars, especially for the stars in the red giant phase. 
However,\cite{2014ApJ...794....3V} find that planet engulfment along the red giant branch is not very sensitive to the stellar mass or mass-loss rates,
but quite sensitive to the planetary mass.

Finally, it should be noted that it is well known that the planet occurrence rate, in addition to the metal content of the star,
does also rise monotonically with the stellar mass \citep[e.g.][]{2010PASP..122..905J}. However, for all the samples studied here (pre-MS, MS, and red giants)
we find a rather flat distribution without any clear mass-planet dependency.

% -------------------------------------------------
\subsection{Can we discard disc instability?}
% -------------------------------------------------

Another possibility is that intermediate-mass stars represent a different stellar population in which
a metal-rich environment is not required for planet formation.
This would naturally explain the lack of a statistically well-founded giant-planet metallicity
correlation in intermediate-mass pre-MS and MS stars.
However, we find that giant stars do show the planet-metallicity correlation.

Thus, it is likely that both formation mechanisms, core-accretion and disc instability might simultaneously
operate in intermediate-mass stars.
The disc mass scales with the stellar mass over several orders of magnitude 
\citep[see][and references therein]{2012A&A...543A..59M,2023ASPC..534..539M}.
%It is expected that high-mass stars should harbour more massive protoplanetary discs \citep{2000prpl.conf..559N}.
%H$_{\alpha}$ measurements in young, low-mass objects suggest that the mass-accretion rate scales
%approximately with the square of the stellar mass \citep[e.g.][]{2011A&A...535A..99M,2012A&A...543A..59M}. This result can be explained assuming that the disc mas
%scales with the mass of the star following the relationship M$_{\rm disc}$ $\propto $ M$_{\star}^{1.2}$ \citep{2011A&A...526A..63A}.
Detailed simulations have confirmed that massive ($>$ 0.1 M$_{\odot}$) protoplanetary discs might cool rapidly enough to become gravitationally unstable. %, and to fragment,
On the other hand, fragmentation appears less likely to produce giant planets around solar-type FGK stars, where the gas cools too slowly for it to fragment into bound clumps
\citep[see][and references therein]{2018exha.book.....P}.

A disc-to-star mass ratio of 0.1 is the usual threshold for triggering gravitational instability \citep{Kratter16}.
However, disc masses inferred from (sub-)mm continuum emission lead to disc-to-star ratios that tend to be an
order of magnitude smaller both for low- and intermediate-mass stars \citep{2012A&A...543A..59M,2021A&A...650A.182G,2022A&A...658A.112S}.
Recently, \citet{2024A&A...686L...1M} argued that gravitational instability is not consistent with the presence of hot Jupiters around intermediate-mass stars, which could be explained via a combination of the core-accretion paradigm and migration.
On the other hand, assuming that disc masses are larger than inferred from (sub-)mm continuum emission, \citet{Dong18} proposed that gravitational instability may explain the multi-arm features commonly observed in discs around Herbig stars. Recently, \citet{Speedie24} inferred a disc-to-star ratio $\sim$ 0.3 for the protoplanet-candidate host AB Aurigae, also providing kinematic evidence of gravitational instability in its disc.

If gravitational instability were the main mechanism for giant-planet formation in intermediate-mass stars,
one would expect a weaker planet-metallicity correlation as the stellar mass increases. 
This is in disagreement with what we found. Thus, even though some specific planet might still be
formed by gravitational instability, our analysis suggests that core-accretion is likely the main formation scenario for gas-giant planets.

%To summarise, two factors might conspire against a robust planet-metallicity in MS intermediate-mass stars.
%To start with, planet formation might lead to an accretion of metal-polluted material on to the surface of the star.
%Secondly, planets might be formed by a non-metallicity dependent mechanism.

%________________________________________________________________
\section{Conclusions}\label{conclusion}
%----------------------------------------------------------------
In this work the metallicity distribution of a large sample
of intermediate-mass stars (1.5 M$_{\odot}$ < M$_{\star}$ < 3.5 M$_{\odot}$) 
at different evolutionary stages is presented with the aim to test the planet-metallicity correlation.
We compare the metallicity distribution of stars with and without known gas-giant companions and perform
a detailed statistical analysis to quantify the strength of the correlations. 

Under the assumption that the gap inferred in 
the protoplanetary disc of group I stars is related to the formation
of giant-planets, we show that intermediate-mass pre-MS do not follow the well
known gas-giant planet-metallicity correlation.
For MS intermediate-mass stars, a tendency of higher metallicity in planet hosts with respect to
its comparison sample is found, but it lacks of statistical support and the strength of the correlation
is significantly lower than for FGK MS stars. Intermediate-mass stars in the red giant branch follow a strong-planet metallicity correlation.

By taking into account the internal evolution of the host stars, we show that the previous results are generally more compatible with the core-accretion,
scenario of planet formation than with the disc instability scenario.
Intermediate-mass pre-MS stars are born fully convective, but they develop a radiative envelope in short-time timescales.
Forming planets in group I sources trap the metal-rich material, and since these stars have radiative envelopes,
metallicity reflects the composition of the stellar surface, polluted by the accreted metal-poor material.
Pre-MS Herbig stars evolve into early-type MS stars which remain radiative during their MS evolution.
Slow non-convective processes might partially mix the metal-poor accreted material, thus preventing us from
finding a statistically significant planet-metallicity correlation. 
When early-type stars leave the MS and evolve along the red giant branch they become convective for the most part
and the gas-giant planet-metallicity correlation is recovered. 

Evolutionary models predict that massive pre-MS stars turn on radiative more rapidly.
In addition, massive giants remain longer radiative during their evolution.
Therefore, more massive stars should be more affected by the accretion of metal-poor material.
However, within our sample, we find that the more massive stars %pre-MS and giant stars
show a stronger planet-metallicity correlation than the less massive stars.
We discuss several possible explanations.
In particular, we argue that the planet engulfment scenario may play a role in explaining the previous result, but it deserves a more careful analysis.

%To summarise our results,
%two factors might conspire against 
%a robust planet-metallicity in MS intermediate-mass stars.
%To start with, planet formation might lead to an accretion of metal-polluted material on to the surface of the star.
%Secondly, planets might still be formed by a non-metallicity dependent mechanism.

A better understanding of the planet-metallicity correlation in intermediate-massive stars will
require the detection of larger samples of known planets as well as accurate masses and radii of the
host stars. 
The forthcoming mission PLATO \citep{2014ExA....38..249R} is expected to make systematic use of asteroseismology to characterise 
planet host stars allowing us to link planetary and stellar evolution. Science targets for PLATO
will include massive stars, red giants, asymptotic giant branch stars and supergiants, as well as white dwarfs.
These, together with detailed models of diffusion, rotation, and other stellar mixing mechanisms
that control how the accreted material onto the stellar surface interchanges with the interior of the star,
will help us to unravel the link between stellar abundance patterns and planet formation.
%intermediate-mass stars. 

%<<<<<<<<<<<<<<<<<<<<<<<<<<<<<<<<<<<<<<<<<<<<<<<<<<<<<<<<<<<<<<<<<<<<
%                      Bibliografia
%>>>>>>>>>>>>>>>>>>>>>>>>>>>>>>>>>>>>>>>>>>>>>>>>>>>>>>>>>>>>>>>>>>>>
%
\bibliographystyle{aa}
\bibliography{pmc_evol.bib}

%<<<<<<<<<<<<<<<<<<<<<<<<<<<<<<<<<<<<<<<<<<<<<<<<<<<<<<<<<<<<<<<<<<<<
%                      Acknowledgements
%>>>>>>>>>>>>>>>>>>>>>>>>>>>>>>>>>>>>>>>>>>>>>>>>>>>>>>>>>>>>>>>>>>>>

\begin{acknowledgements}
J. M. acknowledges support from the Italian Ministero dell'Università e della Ricerca and
the European Union - Next Generation EU through project PRIN 2022 PM4JLH ``Know your little neighbours: characterising low-mass stars and planets in the Solar neighbourhood''.
G. M. M. acknowledges support from the project AST22\_00001\_8 of the Junta de Andaluc\'ia and the
Spanish Ministerio de Ciencia, Innovaci\'on y Universidades funded by the Next Generation EU and the Plan de Recuperaci\'on, Transformaci\'on y
Resiliencia.
I. M. 's research is funded by grants PID2022-138366NA-I00, by the Spanish Ministry of Science and Innovation/State Agency of Research MCIN/AEI/10.13039/501100011033 and by the European Union, and by
a Ram\'on y Cajal fellowship RyC2019-026992-I.
E. V. , B. M. and J. L. G. M. acknowledge the funding by grant  PID2021-127289-NB-I00 from MCIN/AEI/10.13039/501100011033/ and FEDER.

\end{acknowledgements}

%%<<<<<<<<<<<<<<<<<<<<<<<<<<<<<<<<<<<<<<<<<<<<<<<<<<<<<<<<<<<<<<<<<<<<
%%                      Apendix
%%>>>>>>>>>>>>>>>>>>>>>>>>>>>>>>>>>>>>>>>>>>>>>>>>>>>>>>>>>>>>>>>>>>>>
\begin{appendix} %\label{app_tables}
\section{Online material}
\label{app_tables}

%%<<<<<<<<<<<<<<<<<<<<<<<<<<<<<<<<<<<<<<<<<<<<<<<<<<<<<<<<<<<<<<<<<<<<
%% Figure: corner plot of Bayesian fit                     
%%>>>>>>>>>>>>>>>>>>>>>>>>>>>>>>>>>>>>>>>>>>>>>>>>>>>>>>>>>>>>>>>>>>>>
\begin{figure*}[]
\centering
\begin{minipage}{0.33\linewidth}
\includegraphics[scale=0.40]{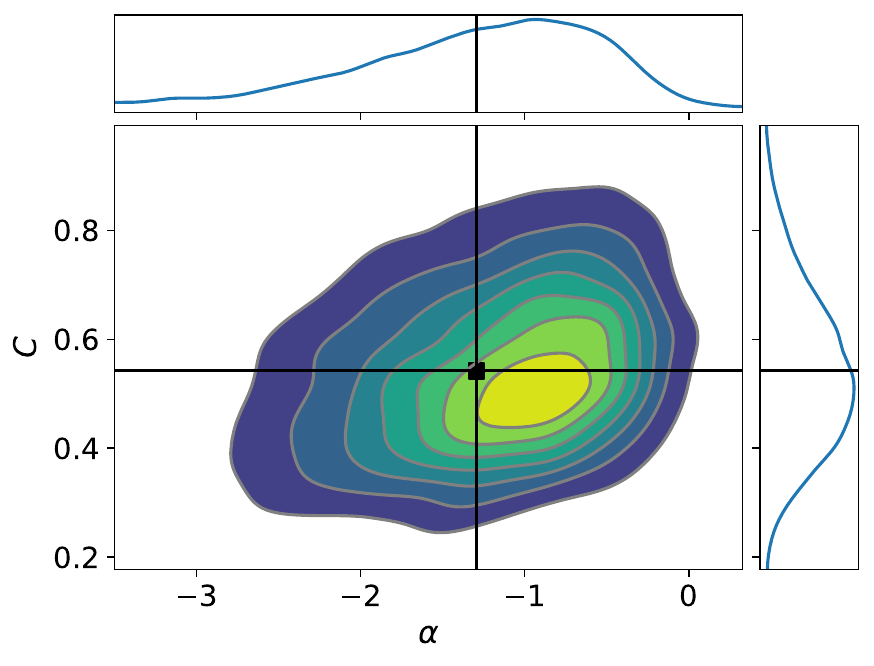}
\end{minipage}
\begin{minipage}{0.33\linewidth}
\includegraphics[scale=0.40]{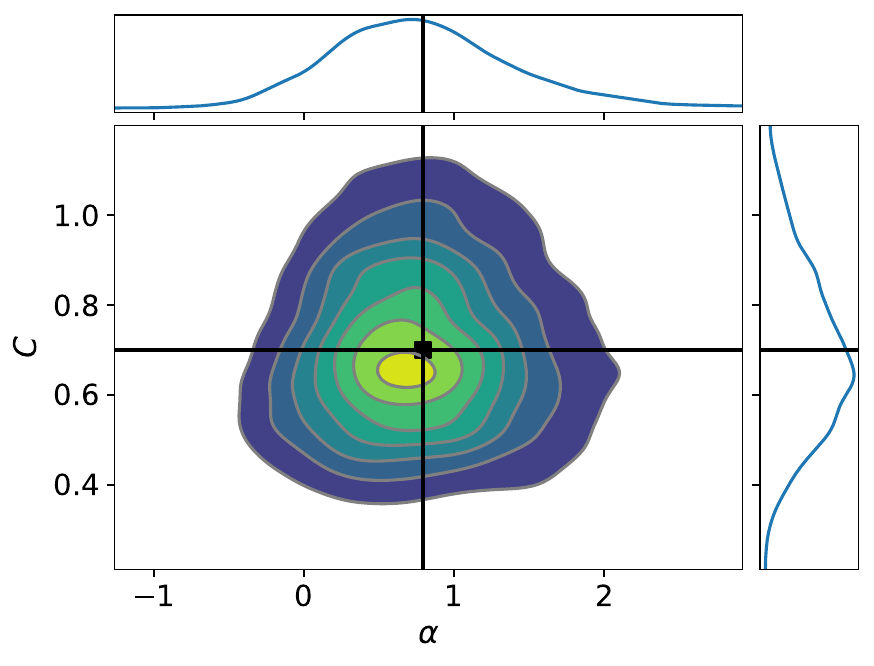}
\end{minipage}
\begin{minipage}{0.33\linewidth}
\includegraphics[scale=0.40]{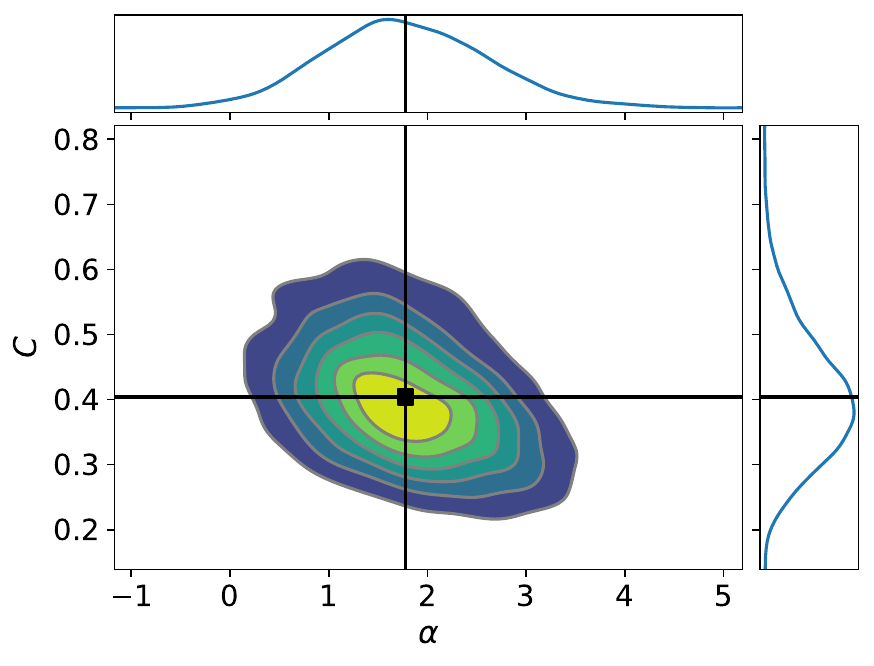} %MESA_evol_giants.pdf}
\end{minipage}
\caption{
Marginal posterior probability distribution functions of the Bayesian fit
to Eq.~\ref{metaleq} for pre-MS stars (left), MS (middle), and giant stars (right).
The vertical line indicates the mean of the distribution.
}
\label{corner_plots}
\end{figure*}

%%<<<<<<<<<<<<<<<<<<<<<<<<<<<<<<<<<<<<<<<<<<<<<<<<<<<<<<<<<<<<<<<<<<<<
%% Figure: Tables
%%>>>>>>>>>>>>>>>>>>>>>>>>>>>>>>>>>>>>>>>>>>>>>>>>>>>>>>>>>>>>>>>>>>>>
%\include{tables/app_table_preMS.tex} 
%\include{tables/app_table_MS.tex}
%\include{tables/app_table_GIANTS.tex}
\begin{table*}
\centering
\caption{Stellar parameters of the pre-MS stars.}
\label{preMS_table}
\begin{tabular}{lccccc}
\hline
Star & Group & T$_{\rm eff}$ & [Fe/H] & M$_{\star}$   & Age   \\
     &       &      (K)      & (dex)  & (M$_{\odot}$) & (Myr) \\
\hline
  HD 9672 	&	II	&	9000	$\pm$	125	&	0.20	$\pm$	0.10	&	1.93	$\pm$	0.03	&	11.0	$\pm$	5.10	\\
  HD 31648 	&	II	&	8000	$\pm$	125	&	-0.25	$\pm$	0.13	&	1.85	$\pm$	0.03	&	7.7	$\pm$	0.30	\\
  UX Ori 	&	II	&	8500	$\pm$	250	&	0.00	$\pm$	0.10	&	1.91	$\pm$	0.02	&	9.8	$\pm$	0.10	\\
  HD 290380 	&	II	&	6250	$\pm$	125	&	0.00	$\pm$	0.10	&	1.59	$\pm$	0.06	&	9.3	$\pm$	1.00	\\
  HD 287823 	&	I	&	8375	$\pm$	125	&	-0.50	$\pm$	0.14	&	1.83	$\pm$	0.04	&	10.6	$\pm$	2.00	\\
  V346 Ori 	&	I	&	7750	$\pm$	250	&	0.00	$\pm$	0.13	&	1.65	$\pm$	0.04	&	16.2	$\pm$	4.40	\\
  CO Ori 	&	II	&	6500	$\pm$	215	&	0.15	$\pm$	0.10	&	2.30	$\pm$	0.33	&	3.9	$\pm$	1.60	\\
  HD 290500 	&	I	&	9500	$\pm$	500	&	0.00	$\pm$	0.10	&	1.85	$\pm$	0.03	&	< 19.9			\\
  HD 244314 	&	II	&	8500	$\pm$	250	&	0.00	$\pm$	0.10	&	2.12	$\pm$	0.06	&	7.0	$\pm$	0.30	\\
  HD 244604 	&	II	&	9000	$\pm$	250	&	0.50	$\pm$	0.10	&	2.16	$\pm$	0.03	&	5.1	$\pm$	0.10	\\
  RY Ori 	&	II	&	6250	$\pm$	194	&	0.00	$\pm$	0.14	&	1.58	$\pm$	0.17	&	9.4	$\pm$	2.30	\\
  HD 245185 	&	I	&	10000	$\pm$	500	&	-0.50	$\pm$	0.30	&	2.24	$\pm$	0.03	&	7.1	$\pm$	0.00	\\
  NV Ori 	&	I	&	7000	$\pm$	125	&	0.14	$\pm$	0.10	&	2.09	$\pm$	0.08	&	5.0	$\pm$	0.50	\\
  HD 37258 	&	II	&	9750	$\pm$	500	&	0.35	$\pm$	0.10	&	2.27	$\pm$	0.06	&	5.9	$\pm$	0.90	\\
  HD 290770 	&	II	&	10500	$\pm$	250	&	-1.00	$\pm$	0.33	&	2.64	$\pm$	0.09	&	4.0	$\pm$	0.20	\\
  BF Ori 	&	II	&	9000	$\pm$	250	&	0.20	$\pm$	0.13	&	1.85	$\pm$	0.08	&	17.1	$\pm$	1.40	\\
  HD 37357 	&	II	&	9500	$\pm$	250	&	0.50	$\pm$	0.20	&	2.80	$\pm$	0.20	&	3.0	$\pm$	0.60	\\
  HD 290764 	&	I	&	7875	$\pm$	375	&	-0.15	$\pm$	0.10	&	1.99	$\pm$	0.04	&	6.1	$\pm$	0.40	\\
  V599 Ori 	&	I	&	8000	$\pm$	250	&	-0.35	$\pm$	0.10	&	2.15	$\pm$	0.11	&	5.1	$\pm$	0.70	\\
  V350 Ori 	&	II	&	9000	$\pm$	250	&	0.50	$\pm$	0.10	&	 < 1.91 		&	< 15.0			\\
  HD 39014 	&	II	&	8000	$\pm$	125	&	0.20	$\pm$	0.20	&	2.48	$\pm$	0.07	&	3.5	$\pm$	0.30	\\
  PDS 124 	&	I	&	10250	$\pm$	250	&	0.10	$\pm$	0.15	&	2.38	$\pm$	0.04	&	6.0	$\pm$	0.40	\\
  LkHa 339 	&	I	&	10500	$\pm$	250	&	0.50	$\pm$	0.12	&	2.52	$\pm$	0.10	&	4.9	$\pm$	0.30	\\
  HBC 217 	&	I	&	6000	$\pm$	125	&	-0.25	$\pm$	0.10	&	1.75	$\pm$	0.10	&	6.9	$\pm$	1.30	\\
  HD 68695 	&	I	&	9250	$\pm$	250	&	-0.50	$\pm$	0.10	&	2.08	$\pm$	0.05	&	8.0	$\pm$	0.80	\\
  GSC 8581-2002 &	I	&	9750	$\pm$	250	&	0.00	$\pm$	0.10	&	2.40	$\pm$	0.05	&	5.1	$\pm$	0.50	\\
  PDS 33 	&	I	&	9750	$\pm$	250	&	-0.50	$\pm$	0.17	&	 < 2.10 		&	< 9.20			\\
  HD 97048 	&	I	&	10500	$\pm$	500	&	-0.50	$\pm$	0.13	&	2.80	$\pm$	0.03	&	3.9	$\pm$	0.30	\\
  HD 100453 	&	I	&	7250	$\pm$	250	&	-0.10	$\pm$	0.10	&	1.60	$\pm$	0.05	&	19.3	$\pm$	0.70	\\
  HD 132947 	&	II	&	10250	$\pm$	250	&	0.20	$\pm$	0.10	&	2.77	$\pm$	0.03	&	3.7	$\pm$	0.40	\\
  HD 135344B 	&	I	&	6375	$\pm$	125	&	0.00	$\pm$	0.13	&	1.46	$\pm$	0.06	&	10.5	$\pm$	0.70	\\
  HD 139614 	&	I	&	7750	$\pm$	250	&	-0.25	$\pm$	0.10	&	1.60	$\pm$	0.01	&	19.3	$\pm$	0.30	\\
  HD 141569 	&	II	&	9500	$\pm$	250	&	-0.50	$\pm$	0.38	&	2.12	$\pm$	0.03	&	8.0	$\pm$	0.00	\\
  HD 142666 	&	II	&	7250	$\pm$	250	&	0.00	$\pm$	0.10	&	1.75	$\pm$	0.02	&	8.7	$\pm$	0.40	\\
  HD 142527 	&	I	&	6500	$\pm$	250	&	0.08	$\pm$	0.13	&	2.20	$\pm$	0.05	&	4.4	$\pm$	0.40	\\
  HD 144432 	&	II	&	7500	$\pm$	250	&	0.15	$\pm$	0.10	&	1.82	$\pm$	0.02	&	8.0	$\pm$	0.10	\\
  V718 Sco 	&	II	&	7750	$\pm$	250	&	0.00	$\pm$	0.10	&	1.71	$\pm$	0.03	&	9.0	$\pm$	0.30	\\
  HD 149914 	&	II	&	9500	$\pm$	125	&	0.00	$\pm$	0.14	&	3.07	$\pm$	0.06	&	2.0	$\pm$	0.00	\\
  HD 150193 	&	II	&	9250	$\pm$	250	&	0.20	$\pm$	0.10	&	2.25	$\pm$	0.04	&	6.0	$\pm$	0.50	\\
  HD 163296 	&	II	&	9000	$\pm$	250	&	0.20	$\pm$	0.10	&	1.91	$\pm$	0.06	&	10.0	$\pm$	5.80	\\
  HD 169142 	&	I	&	7250	$\pm$	125	&	-0.50	$\pm$	0.33	&	1.55	$\pm$	0.02	&	< 20			\\
  HD 179218 	&	I	&	9500	$\pm$	250	&	-0.50	$\pm$	0.17	&	2.99	$\pm$	0.03	&	2.3	$\pm$	0.20	\\
  PX Vul 	&	II	&	6500	$\pm$	125	&	0.00	$\pm$	0.10	&	2.59	$\pm$	0.16	&	3.0	$\pm$	0.40	\\
  HD 199603 	&	II	&	7500	$\pm$	125	&	0.00	$\pm$	0.10	&	2.10	$\pm$	0.05	&	5.2	$\pm$	0.40	\\
\hline
\end{tabular}
\end{table*}

\begin{table*}
\centering
\caption{Stellar parameters of the Main-Sequence stars.}
\label{MS_table}
\begin{tabular}{lccccc}
\hline
Star & Planet  & T$_{\rm eff}$ & [Fe/H] & M$_{\star}$   & Age   \\
     &         &      (K)      & (dex)  & (M$_{\odot}$) & (Gyr) \\
\hline
WASP-33		&	y	&	7373	$\pm$	164	&	0.13	$\pm$	0.16	&	1.533	$\pm$	0.067	&	0.527	$\pm$	0.369	\\
HD 23281	&		&	7761	$\pm$	135	&	0.11	$\pm$	0.10	&	1.613	$\pm$	0.056	&	0.387	$\pm$	0.244	\\
HAT-P-70	&	y	&	8450	$\pm$	195	&	-0.29	$\pm$	0.15	&	1.648	$\pm$	0.092	&	0.840	$\pm$	0.338	\\
beta Pic	&	y	&	8084	$\pm$	130	&	-0.26	$\pm$	0.14	&	1.551	$\pm$	0.062	&	0.593	$\pm$	0.332	\\
V435 Car	&		&	7510	$\pm$	165	&	-0.11	$\pm$	0.09	&	1.546	$\pm$	0.078	&	0.900	$\pm$	0.463	\\
HD 50445	&		&	7922	$\pm$	117	&	-0.28	$\pm$	0.16	&	1.560	$\pm$	0.062	&	1.031	$\pm$	0.281	\\
HD 56537	&		&	8231	$\pm$	122	&	-0.43	$\pm$	0.16	&	1.810	$\pm$	0.064	&	1.053	$\pm$	0.090	\\
KELT-19A	&	y	&	7500	$\pm$	130	&	0.41	$\pm$	0.17	&	1.656	$\pm$	0.048	&	0.337	$\pm$	0.210	\\
KELT-17		&	y	&	7471	$\pm$	210	&	0.51	$\pm$	0.17	&	1.703	$\pm$	0.042	&	0.333	$\pm$	0.208	\\
HAT-P-69	&	y	&	7750	$\pm$	245	&	0.07	$\pm$	0.15	&	1.722	$\pm$	0.086	&	0.605	$\pm$	0.358	\\
MASCARA-4	&	y	&	7810	$\pm$	165	&	-0.17	$\pm$	0.12	&	1.618	$\pm$	0.077	&	1.102	$\pm$	0.290	\\
HD 88955	&		&	8733	$\pm$	154	&	-0.45	$\pm$	0.13	&	1.778	$\pm$	0.068	&	0.951	$\pm$	0.129	\\
HD 95086	&	y	&	7593	$\pm$	122	&	-0.05	$\pm$	0.16	&	1.511	$\pm$	0.061	&	0.520	$\pm$	0.345	\\
HR 4502		&		&	9569	$\pm$	253	&	-0.11	$\pm$	0.10	&	2.058	$\pm$	0.079	&	0.261	$\pm$	0.125	\\
HD 105850	&		&	9052	$\pm$	167	&	-0.24	$\pm$	0.13	&	1.883	$\pm$	0.081	&	0.599	$\pm$	0.173	\\
WASP-167	&	y	&	6770	$\pm$	210	&	0.03	$\pm$	0.15	&	1.581	$\pm$	0.058	&	1.504	$\pm$	0.406	\\
HD 115820	&		&	7610	$\pm$	135	&	-0.02	$\pm$	0.09	&	1.520	$\pm$	0.061	&	0.503	$\pm$	0.334	\\
HD 129926	&		&	7101	$\pm$	167	&	0.38	$\pm$	0.12	&	1.588	$\pm$	0.067	&	0.512	$\pm$	0.371	\\
WASP-189	&	y	&	7946	$\pm$	136	&	0.05	$\pm$	0.13	&	1.866	$\pm$	0.073	&	0.814	$\pm$	0.177	\\
HD 146624	&		&	9489	$\pm$	184	&	0.01	$\pm$	0.11	&	2.006	$\pm$	0.053	&	0.208	$\pm$	0.088	\\
HD 153053	&		&	7916	$\pm$	129	&	-0.22	$\pm$	0.15	&	1.619	$\pm$	0.068	&	1.099	$\pm$	0.238	\\
HD 159492	&		&	8076	$\pm$	128	&	-0.02	$\pm$	0.11	&	1.707	$\pm$	0.069	&	0.468	$\pm$	0.259	\\
Vega		&		&	9505	$\pm$	188	&	-0.72	$\pm$	0.08	&	2.037	$\pm$	0.075	&	0.721	$\pm$	0.071	\\
KELT-20		&	y	&	8652	$\pm$	160	&	-0.30	$\pm$	0.15	&	1.703	$\pm$	0.078	&	0.709	$\pm$	0.258	\\
HD 188228	&		&	10262	$\pm$	172	&	0.02	$\pm$	0.08	&	2.201	$\pm$	0.037	&	0.144	$\pm$	0.039	\\
KELT-9		&	y	&	9329	$\pm$	118	&	-0.08	$\pm$	0.14	&	2.215	$\pm$	0.070	&	0.482	$\pm$	0.078	\\
MASCARA-1	&	y	&	7687	$\pm$	238	&	-0.10	$\pm$	0.13	&	1.704	$\pm$	0.074	&	0.988	$\pm$	0.287	\\
Fomalhaut	&		&	8745	$\pm$	195	&	0.11	$\pm$	0.15	&	1.949	$\pm$	0.071	&	0.295	$\pm$	0.154	\\
\hline
\end{tabular}
\end{table*}

\begin{table*}
\centering
\caption{Stellar parameters of the red giant stars.}
\label{giants_table}
\begin{tabular}{lccccc}
\hline
Star & Planet  & T$_{\rm eff}$ & [Fe/H] & M$_{\star}$   & Age   \\
     &         &      (K)      & (dex)  & (M$_{\odot}$) & (Gyr) \\
\hline
HIP 671	        &	y	&	4829	$\pm$	10	&	-0.13	$\pm$	0.02	&	1.68	$\pm$	0.05	&	1.97	$\pm$	0.17	\\
HIP    729	&		&	4811	$\pm$	23	&	0.04	$\pm$	0.04	&	1.81	$\pm$	0.08	&	1.77	$\pm$	0.19	\\
HIP    873	&		&	4849	$\pm$	15	&	0.03	$\pm$	0.02	&	1.78	$\pm$	0.05	&	1.72	$\pm$	0.14	\\
HIP 5364	&	y	&	4594	$\pm$	25	&	0.10	$\pm$	0.04	&	1.63	$\pm$	0.17	&	2.45	$\pm$	0.82	\\
HIP   6999	&		&	4961	$\pm$	20	&	0.07	$\pm$	0.03	&	2.43	$\pm$	0.08	&	0.76	$\pm$	0.09	\\
HIP   7097	&		&	4936	$\pm$	40	&	-0.09	$\pm$	0.06	&	3.62	$\pm$	0.14	&	0.24	$\pm$	0.03	\\
HIP   7607	&		&	4396	$\pm$	38	&	0.11	$\pm$	0.07	&	1.85	$\pm$	0.20	&	1.78	$\pm$	0.49	\\
HIP   7719	&		&	5097	$\pm$	20	&	-0.12	$\pm$	0.02	&	2.32	$\pm$	0.10	&	0.77	$\pm$	0.11	\\
HIP 8928	&	y	&	4917	$\pm$	10	&	-0.25	$\pm$	0.02	&	1.65	$\pm$	0.08	&	1.97	$\pm$	0.24	\\
HIP   9222	&		&	4803	$\pm$	10	&	-0.15	$\pm$	0.02	&	1.52	$\pm$	0.04	&	2.51	$\pm$	0.20	\\
HIP 11791	&	y	&	4884	$\pm$	15	&	-0.01	$\pm$	0.02	&	1.94	$\pm$	0.09	&	1.54	$\pm$	0.19	\\
HIP  12247	&	y	&	4864	$\pm$	23	&	-0.03	$\pm$	0.03	&	1.83	$\pm$	0.09	&	1.68	$\pm$	0.18	\\
HIP  13531	&		&	5025	$\pm$	25	&	-0.20	$\pm$	0.03	&	2.91	$\pm$	0.08	&	0.43	$\pm$	0.04	\\
HIP  19038	&		&	4808	$\pm$	28	&	0.09	$\pm$	0.04	&	2.24	$\pm$	0.12	&	1.09	$\pm$	0.21	\\
HIP  20889	&	y	&	4911	$\pm$	25	&	0.15	$\pm$	0.04	&	2.63	$\pm$	0.07	&	0.63	$\pm$	0.07	\\
HIP 31674	&	y	&	4884	$\pm$	10	&	-0.06	$\pm$	0.02	&	1.71	$\pm$	0.03	&	1.83	$\pm$	0.09	\\
HIP  36616	&	y	&	4669	$\pm$	20	&	0.13	$\pm$	0.04	&	1.98	$\pm$	0.19	&	1.48	$\pm$	0.46	\\
HIP  37826	&	y	&	4887	$\pm$	18	&	0.09	$\pm$	0.03	&	2.03	$\pm$	0.07	&	1.31	$\pm$	0.17	\\
HIP  42528	&		&	5128	$\pm$	20	&	0.01	$\pm$	0.03	&	1.76	$\pm$	0.05	&	1.75	$\pm$	0.13	\\
HIP  57820	&	y	&	4978	$\pm$	30	&	0.14	$\pm$	0.04	&	1.56	$\pm$	0.09	&	2.68	$\pm$	0.47	\\
HIP  59646	&		&	5132	$\pm$	23	&	0.10	$\pm$	0.03	&	1.52	$\pm$	0.03	&	2.87	$\pm$	0.14	\\
HIP  59847	&		&	4872	$\pm$	20	&	-0.20	$\pm$	0.03	&	1.61	$\pm$	0.12	&	2.16	$\pm$	0.44	\\
HIP  59856	&		&	4581	$\pm$	28	&	-0.12	$\pm$	0.05	&	1.85	$\pm$	0.14	&	1.57	$\pm$	0.32	\\
HIP  61740	&	y	&	4605	$\pm$	40	&	0.35	$\pm$	0.06	&	2.94	$\pm$	0.10	&	0.46	$\pm$	0.07	\\
HIP  69612	&		&	4809	$\pm$	23	&	-0.08	$\pm$	0.03	&	1.61	$\pm$	0.11	&	2.29	$\pm$	0.44	\\
HIP  70038	&		&	5071	$\pm$	23	&	-0.04	$\pm$	0.03	&	2.04	$\pm$	0.07	&	1.11	$\pm$	0.10	\\
HIP  74793	&	y	&	4260	$\pm$	33	&	0.00	$\pm$	0.06	&	1.69	$\pm$	0.18	&	2.08	$\pm$	0.65	\\
HIP  76311	&	y	&	4601	$\pm$	28	&	0.34	$\pm$	0.05	&	2.16	$\pm$	0.18	&	1.19	$\pm$	0.33	\\
HIP  80816	&		&	5000	$\pm$	20	&	-0.10	$\pm$	0.03	&	2.89	$\pm$	0.07	&	0.45	$\pm$	0.04	\\
HIP  88765	&		&	4998	$\pm$	15	&	-0.01	$\pm$	0.03	&	2.68	$\pm$	0.07	&	0.56	$\pm$	0.06	\\
HIP  88836	&		&	4647	$\pm$	33	&	0.21	$\pm$	0.06	&	1.52	$\pm$	0.10	&	2.96	$\pm$	0.56	\\
HIP  89047	&	y	&	4983	$\pm$	10	&	0.03	$\pm$	0.02	&	1.53	$\pm$	0.02	&	2.69	$\pm$	0.09	\\
HIP  89826	&		&	4639	$\pm$	28	&	0.13	$\pm$	0.05	&	2.62	$\pm$	0.18	&	0.66	$\pm$	0.17	\\
HIP  89918	&		&	5061	$\pm$	18	&	-0.09	$\pm$	0.02	&	2.45	$\pm$	0.07	&	0.69	$\pm$	0.08	\\
HIP  91852	&	y	&	4775	$\pm$	30	&	-0.12	$\pm$	0.04	&	1.83	$\pm$	0.24	&	1.62	$\pm$	0.60	\\
HIP  95124	&	y	&	4998	$\pm$	15	&	0.20	$\pm$	0.03	&	1.71	$\pm$	0.07	&	2.07	$\pm$	0.24	\\
HIP  95822	&		&	4819	$\pm$	25	&	0.07	$\pm$	0.03	&	1.92	$\pm$	0.14	&	1.61	$\pm$	0.29	\\
HIP  96016	&		&	4997	$\pm$	13	&	-0.13	$\pm$	0.02	&	1.57	$\pm$	0.04	&	2.27	$\pm$	0.15	\\
HIP  98920	&		&	4775	$\pm$	28	&	0.09	$\pm$	0.04	&	1.60	$\pm$	0.08	&	2.43	$\pm$	0.33	\\
HIP  99841	&		&	5024	$\pm$	15	&	-0.11	$\pm$	0.02	&	2.16	$\pm$	0.09	&	1.00	$\pm$	0.15	\\
HIP  99913	&		&	4805	$\pm$	15	&	0.07	$\pm$	0.03	&	1.87	$\pm$	0.05	&	1.62	$\pm$	0.14	\\
HIP 100541	&		&	4932	$\pm$	20	&	-0.06	$\pm$	0.03	&	1.95	$\pm$	0.10	&	1.44	$\pm$	0.23	\\
HIP 101936	&		&	4796	$\pm$	33	&	0.07	$\pm$	0.05	&	1.83	$\pm$	0.13	&	1.76	$\pm$	0.31	\\
HIP 102532	&		&	4801	$\pm$	18	&	0.13	$\pm$	0.03	&	1.94	$\pm$	0.05	&	1.48	$\pm$	0.12	\\
HIP 103004	&		&	5261	$\pm$	15	&	-0.05	$\pm$	0.02	&	2.40	$\pm$	0.05	&	0.70	$\pm$	0.04	\\
HIP 103519	&		&	4993	$\pm$	20	&	0.12	$\pm$	0.03	&	2.33	$\pm$	0.05	&	0.80	$\pm$	0.04	\\
HIP 103527	&	y	&	5106	$\pm$	30	&	0.07	$\pm$	0.04	&	2.28	$\pm$	0.04	&	0.84	$\pm$	0.03	\\
HIP 104202	&	y	&	5067	$\pm$	20	&	-0.11	$\pm$	0.03	&	1.53	$\pm$	0.05	&	2.50	$\pm$	0.20	\\
HIP 105411	&		&	4809	$\pm$	25	&	-0.10	$\pm$	0.03	&	1.77	$\pm$	0.15	&	1.65	$\pm$	0.43	\\
HIP 106093	&		&	4953	$\pm$	15	&	0.06	$\pm$	0.03	&	2.02	$\pm$	0.07	&	1.26	$\pm$	0.16	\\
HIP 107251	&	y	&	4953	$\pm$	45	&	0.22	$\pm$	0.06	&	1.74	$\pm$	0.21	&	1.97	$\pm$	0.70	\\
HIP 108513	&	y	&	4889	$\pm$	15	&	0.17	$\pm$	0.03	&	1.56	$\pm$	0.04	&	2.72	$\pm$	0.16	\\
HIP 109577	&	y	&	4993	$\pm$	15	&	0.02	$\pm$	0.02	&	1.63	$\pm$	0.03	&	2.24	$\pm$	0.12	\\
HIP 111944	&		&	4352	$\pm$	25	&	-0.19	$\pm$	0.05	&	1.66	$\pm$	0.16	&	1.97	$\pm$	0.57	\\
HIP 112158	&		&	4986	$\pm$	20	&	-0.21	$\pm$	0.03	&	3.30	$\pm$	0.09	&	0.30	$\pm$	0.03	\\
HIP 112242	&		&	5035	$\pm$	20	&	0.10	$\pm$	0.03	&	2.59	$\pm$	0.04	&	0.59	$\pm$	0.02	\\
HIP 115830	&		&	4756	$\pm$	25	&	0.08	$\pm$	0.04	&	1.83	$\pm$	0.05	&	1.72	$\pm$	0.13	\\
HIP 115919	&		&	5034	$\pm$	15	&	0.05	$\pm$	0.03	&	2.39	$\pm$	0.06	&	0.74	$\pm$	0.05	\\
HIP 117375	&		&	5010	$\pm$	25	&	-0.05	$\pm$	0.03	&	2.32	$\pm$	0.09	&	0.85	$\pm$	0.12	\\
\hline
\end{tabular}
\end{table*}

\end{appendix} 
%++++++++++++++++++++++++
\end{document}